%% file: main.tex
\documentclass[acmsmall,screen]{acmart}
\usepackage{xspace}
\usepackage{enumitem}  
\usepackage{float}
\usepackage[ruled,vlined,linesnumbered]{algorithm2e}
\usepackage{float}
\usepackage{booktabs}
\usepackage{multirow}
 \usepackage[dvipsnames,table]{xcolor}
\usepackage{array}
\usepackage{graphicx} 
\usepackage{wrapfig}  

\definecolor{ResolveBG}{RGB}{255,255,204}
\newcolumntype{Y}{>{\columncolor{ResolveBG}}c}
\definecolor{ForestGreen}{RGB}{34,139,34} 
\definecolor{BrickRed}{RGB}{203,65,84}    
\newcommand{\good}[1]{\textcolor{ForestGreen}{\bfseries #1}}
\newcommand{\bad}[1]{\textcolor{BrickRed}{#1}}

\RestyleAlgo{ruled}
\SetAlgoInsideSkip{smallskip}
\SetAlgoNlRelativeSize{-1}
\include{tool}
\AtBeginDocument{%
  }

\setcopyright{acmlicensed}
\copyrightyear{2018}
\acmYear{2018}
\acmDOI{XXXXXXX.XXXXXXX}
\acmConference[Conference acronym 'XX]{Make sure to enter the correct
  conference title from your rights confirmation email}{June 03--05,
  2018}{Woodstock, NY}
\acmISBN{978-1-4503-XXXX-X/2018/06}

\begin{document}

\title{SWE Data Construction, Automatically!}

\author{Lianghong Guo}
\email{guolh8@mail2.sysu.edu.cn}
\affiliation{%
  \institution{Sun Yat-sen University}
  \city{Zhuhai City}
  \country{China}
}
\author{Yanlin Wang}
\authornote{Corresponding author.}
\email{yanlin-wang@outlook.com}
\affiliation{%
  \institution{Sun Yat-sen University}
  \city{Zhuhai City}
  \country{China}
}
\author{Caihua Li}
\email{1471890903@qq.com}
\affiliation{%
  \institution{Sun Yat-sen University}
  \city{Zhuhai City}
  \country{China}
}
\author{Wei Tao}
\authornotemark[1]
\email{wtao@ieee.org}
\affiliation{%
  \institution{Independent Researcher}
  \city{Shenzhen City}
  \country{China}
}
\author{Pengyu Yang}
\email{2454700600@qq.com}
\affiliation{%
  \institution{Sun Yat-sen University}
  \city{Zhuhai City}
  \country{China}
}

\author{Jiachi Chen}
\email{chenjch86@mail.sysu.edu.cn}
\affiliation{%
  \institution{Sun Yat-sen University}
  \city{Zhuhai City}
  \country{China}
}

\author{Haoyu Song}
\email{songhaoyu1@huawei.com}
\affiliation{%
  \institution{Huawei Technologies Co, Ltd}
  \city{Beijing City}
  \country{China}
}
\author{Duyu Tang}
\email{tangduyu@huawei.com}
\affiliation{%
  \institution{Huawei Technologies Co, Ltd}
  \city{Beijing City}
  \country{China}
}
\author{Zibin Zheng}
\email{zhzibin@mail.sysu.edu.cn}
\affiliation{%
  \institution{Sun Yat-sen University}
  \city{Zhuhai City}
  \country{China}
}


\renewcommand{\shortauthors}{Trovato et al.}
\acmArticleType{Review}

\begin{abstract}
Constructing large-scale datasets for the GitHub issue resolution task is crucial for both training and evaluating the software engineering capabilities of Large Language Models (LLMs). However, the existing GitHub issue resolution data construction pipeline is challenging and labor-intensive. We identify three key limitations in existing pipelines: (1) test patches collected 
often omit binary file changes; (2) the manual construction of evaluation environments is labor-intensive; and (3) the fail2pass validation phase requires manual inspection of test logs and writing custom parsing code to extract test status from logs.
In this paper, we propose \method, a fully automated issue resolution data construction
pipeline, to resolve these limitations. First, our pipeline automatically recovers missing binary test files and ensures the correctness of test patches. Second, we introduce \setupagent, a LLM-based multi-agent system that automates evaluation environment construction. Third, we introduce a standardized, exit-code-based log parsing method to automatically extract test status, enabling a fully automated fail2pass validation. 
Experiments on 671 real-world GitHub issues across four programming languages show that our method can effectively construct valid evaluation environments for GitHub issues at a reasonable cost. For example, with GPT-4.1 mini, our \setupagent constructs 337 valid task instances out of 671 issues, at \$0.047 per instance.
Our ablation study further shows the effectiveness of different components of \setupagent. We also demonstrate through manual inspection that our exit-code-based fail2pass validation method is highly accurate, achieving an F1 score of 0.99. Additionally, we conduct an exploratory experiment to investigate whether we can use \method to enhance models' software engineering ability. After training five models on 2,809 Python task instances collected by our method, all models show improved software engineering ability. For example, the resolve rate of a trained Qwen2.5-Coder-14B-Instruct on SWE-bench Verified increases from 5.8\% to 21.0\%. We hope our method can accelerate the construction of large-scale, high-quality GitHub issue resolution datasets for both training and evaluation.     
\end{abstract}




\received{20 February 2007}
\received[revised]{12 March 2009}
\received[accepted]{5 June 2009}

\maketitle

\section{Introduction}

The GitHub issue resolution task, which involves addressing real-world software issues like bug fixing and feature enhancements~\cite{BissyandeLJRKT13,TaoZWZWZ24}, is a crucial aspect of software maintenance~\cite{jimenez2023swe,guo2025omnigirl}. Given its practical importance, the task has become a key benchmark for evaluating the software engineering capabilities of Large Language Models (LLMs)~\cite{xia2024agentless,liu2024marscode,tao2024magis,yang2024swem,chen2024coder,zhang2024autocoderover,arora2024masai,wang2024opendevin,ma2024understand}. A prominent example is SWE-bench~\cite{jimenez2023swe}, a large-scale benchmark for evaluation, whose success has inspired many subsequent benchmarks that extend coverage to more languages and issue types~\cite{zan2024swe,guo2025omnigirl,aleithan2024swe,yang2024swem,zan2025multi,he2025swe}. More recently, 
researchers have begun to construct training datasets to improve the software engineering ability of models ~\cite{pan2024training,yang2025swe,jain2025r2e}. For example, SWE-Gym~\cite{pan2024training} collects a large-scale GitHub issue resolution dataset from Python repositories for agent-based model training, and achieves significant performance improvements on this task.

\begin{figure}[h]
    \centering
    \includegraphics[width=\linewidth]{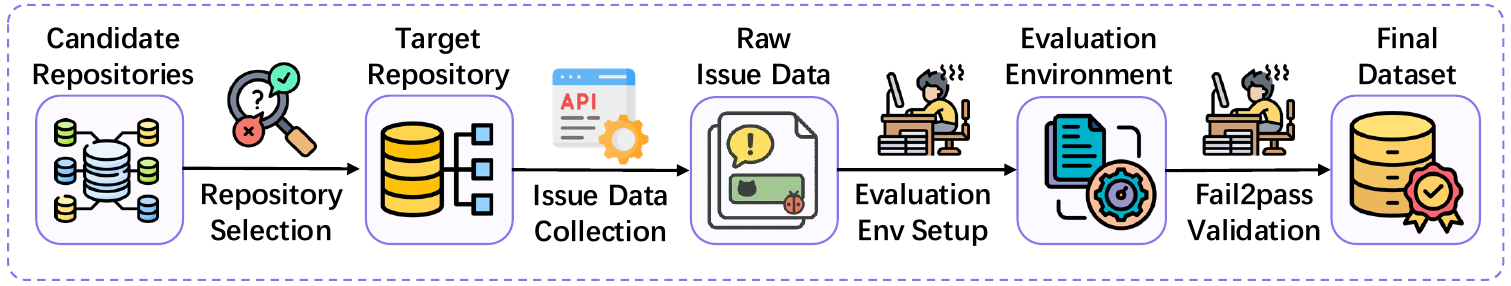}
    \caption{Traditional pipeline of GitHub issue resolution data construction.}
    \label{fig:pipeline}
\end{figure}

Existing works~\cite{zan2024swe,yang2025swe,zhang2025swe,yang2024swem,jimenez2023swe} typically construct GitHub issue resolution datasets following the pipeline of SWE-bench~\cite{jimenez2023swe}. As shown in Figure~\ref{fig:pipeline}, this pipeline consists of four stages. First, in the Repository Selection stage, target repositories are selected, typically based on popularity metrics like GitHub stars. Next, during Issue Data Collection, predefined patterns and the GitHub API are used to collect pairs of issues and their related pull requests as raw task instances. In the next stage, an evaluation environment is constructed for each task instance, consisting of a runtime environment and a script for running the relevant tests.   Finally, the Fail2pass Validation stage validates each instance. Following the method from SWE-bench, tests are run both before and after applying the ground truth patch, and an instance is retained only if the test status is ``fail'' before and ``pass'' after the grount truth patch is applied. While this pipeline is widely adopted, the process is challenging and labor-intensive due to several problems, which we summarize as follows:



\begin{enumerate}[label={\bfseries P\arabic*}]
\item  \textbf{Binary Test Files Missing in Issue Collection Stage.} We find that for task instances where the test patch includes changes to binary test files, the content of these changes is empty when collected via the GitHub API. This causes errors in building the test environment for this instance. This problem is especially prevalent in image processing and visualization repositories, such as Pillow (30.43\% of instances affected) and matplotlib (22.10\%). 
\item \textbf{Manual Evaluation Environment Construction.} The existing pipeline requires significant manual effort to construct an evaluation environment for each task instance. This complexity stems from the diversity of programming languages and repository configurations, resulting in highly varied dependencies and test commands. Furthermore, repositories often have multiple versions, and dependencies or test frameworks can change between them, making the environment setup particularly complex.
\item \textbf{Manual Fail2Pass Validation.} This stage relies on manual effort to inspect test logs and write custom parser code to extract the final test status. The lack of consistency in test log formats complicates this task. Different repositories use different testing commands, resulting in varied log formats. Furthermore, the log format can change as the project evolves, even within the same repository  requiring constant rewriting of parsers.
\end{enumerate}

In this paper, we propose \textbf{\method}, an automatic issue resolution dataset construction pipeline to address these problems.  First, to resolve the binary test file missing issue, we automatically download the missing binary files and then remove their code changes from the test patch, to ensure the correctness of the test environment \textbf{(addressing P1)}. Second, we introduce \textbf{\setupagent}, a multi-agent framework that automates evaluation environment construction. Through the collaboration among different agents, \setupagent automatically collects information from the repository to create an environment construction script and corresponding test scripts. It also automatically executes tests and refines the environment based on the feedback. To further improve efficiency, the framework uses an Evaluation Environment Memory Pool to reuse previously successful setups \textbf{(addressing P2)}. 
Third, we propose an exit-code-based log parsing method to automate the fail2pass validation stage. Inspired by the observation that mainstream test frameworks report test outcomes via exit codes~\cite{wiki_exit_status,junit_exit,mocha_exit,pytest_exit}, we standardize test status collection by capturing the exit code from test commands. Using this method, we can simply extract test status from logs without checking logs and writing parser code manually. \textbf{(addressing P3)}.


\textbf{To investigate the effectiveness of \method, we evaluate our method on 671 issues sampled from 12  repositories across four languages: Python, Java, JavaScript, and TypeScript.} Through these experiments in Section~\ref{sec:eval}, we identify several key findings: (1) First, experiments in Section~\ref{sec:effectiveness} show that \setupagent can effectively construct evaluation environments for issues at a reasonable cost. For example, with GPT-4.1 mini, \setupagent successfully constructs 337 valid task instances out of 671 issues (50.2\%) at an average cost of \$0.047 per instance. 
(2) Second, ablation experiments in Section~\ref{sec:ablation} demonstrate effectiveness of different components of \setupagent.  (3) Third, through manual verification in Section~\ref{sec:rq3}, we demonstrate that our exit-code-based fail2pass validation method is highly accurate, achieving an F1 score of 0.99. Besides, we also verify that the exit code is a reliable indicator to determine the test status.


\textbf{To investigate whether \method can be used to enhance the software engineering ability of models, we conduct an exploratory experiment in Section~\ref{sec:exploratory_exp}.} First, we use our method to collect 2,809 task instances automatically. Second, following previous studies~\cite{pan2024training,jain2025r2e,yang2025swe}, we use \Kimi~\cite{kimi_k2_2025} as the agent model and the DeepSWE agent framework~\cite{luo2025deepswe} to generate 2,809 corresponding agent trajectories. Finally, we finetune five LLMs on the collected trajectories based on the collected task instances. The evaluation results demonstrate that the performance of all models improves after finetuning. For example, the resolve rate of a finetuned Qwen2.5-Coder-14B-Instruct~\cite{qwen_qwen2_5_coder_2024} on SWE-bench Verified~\cite{openai2024swe} increases from 5.8\% to 21.0\%. These findings suggest the potential of \method for enhanching the software engineering ability of models.

Our main contributions are summarized as follows:
\begin{itemize}
  \item We present \method, the first open-source fully automatic pipeline for constructing GitHub issue resolution benchmarks across multiple languages. Our code, datasets, and model weights are released at \url{https://github.com/DeepSoftwareAnalytics/swe-factory}.


  \item We propose \setupagent, a multi-agent framework that automates evaluation environment construction. Experiments show this method can efficiently construct valid evaluation environments for issue across different langauges.

 \item We propose an exit-code-based log parsing method to automate the fail2pass validation stage. Our human evaluation demonstrates the high accuracy of this method.

 \item Our exploratory experiment demonstrates \method can effectively increase the software engineering ability of models. For example, after agent training using 2809 task instances collected by our method, the resolve rate of  Qwen2.5-Coder-14B-Instruct
on SWE-bench Verified increases from 5.8\% to 21.0\%.
\end{itemize}

\section{Background of GitHub Issue Resolution Data Construction Pipeline}

Following the method of SWE-bench~\cite{jimenez2023swe}, existing works~\cite{guo2025omnigirl,zan2024swe,zan2025multi,yang2024swem} often use a four-stage pipeline for constructing GitHub issue resolution datasets, including: (1) Repository Selection, (2) Issue Data Collection, (3) Evaluation Environment Construction, and (4) Task Validation. We introduce these stages in detail as follows.

\textbf{Repository Selection.}  In this stage, researchers decide which GitHub repositories to use for collecting issue resolution data. According to previous studies~\cite{zan2024swe,yang2024swem,jimenez2023swe,yang2025swe,guo2025omnigirl}, researchers often prioritize popular repositories, as measured by GitHub stars or download counts (e.g., via pip). These repositories generally feature a higher volume of issue-resolving activity, leading to a larger and more diverse collection of issue data, and possess comprehensive contribution documentation, which is beneficial for constructing the evaluation environment.

\textbf{Issue Data Collection.} This stage aims to collect issue-pull request pairs, which constitute the task instances for the GitHub issue resolution task. Existing works typically use the GitHub API and predefined patterns to collect issues and their corresponding pull requests. In these pairs, the issue's description serves as the task input. The pull request provides two key components: a ``test patch'' containing test cases to verify the resolution, and a ``gold patch'' that represents the ground-truth solution. A common filtering step is to discard pairs where the pull request does not include changes to test files. Additionally, metadata such as the base commit hash is collected to allow the repository to be checked out to the state it was in before the fix was applied. While this collection process is largely automated, we have identified a critical problem: for issues involving binary test files, the binary files retrieved via the GitHub API are often empty. This leads to errors during the evaluation environment construction, which we discuss further in Section~\ref{sec:binary_files_missing}.

\textbf{Evaluation Environment Construction.} This stage is aimed at building a runnable environment and preparing test scripts for each task instance collected in the previous step. To construct the runtime environment, the process begins by cloning the issue's repository and using the collected base commit to revert it to the state before the fix was applied. Next, all necessary software dependencies are installed to ensure the code can be executed properly. For the testing script, the test patch is first applied to introduce the tests that verify the issue's resolution. Then, the specific command required to run these target tests is created. This entire process is highly dependent on manual effort, as it often requires researchers to consult documentation for different repositories and their various versions to install the correct dependencies and determine the right test commands, making it a time-consuming phase. In our approach, we automate this process by generating a Dockerfile to define the runtime environment and a bash script to run the tests. We accomplish this generation using an LLM-based multi-agent system, introduced in Section~\ref{sec:agent_roles}.

\textbf{Fail2pass Validation.} This stage aims to filter the collected task instances to retain only the valid ones. Following the rules from SWE-bench, a task is considered valid based on a two-step testing process. First, the tests for the issue are run before applying the gold patch, and they must fail. Second, the same tests are run again after applying the gold patch, and this time they must pass. This "fail-to-pass" change shows that the patch correctly fixes the issue. The most time-consuming part of this stage is getting the test results from the logs. This is because researchers need to carefully check many different log file styles and write specific code to parse them. In Section~\ref{sec:grading}, we introduce our method to automate this task validation step.

\section{Methodology}

In this section, we introduce our framework, \method, designed to automate the GitHub issue resolution data construction pipeline. Our framework enhances this process across three main stages. First, for issue collection, we identify and resolve an issue within the SWE-bench script where it failed to collect certain binary test resources (Section~\ref{sec:binary_files_missing}). Second, to automate environment setup, we introduce \setupagent, a multi-agent system detailed in Sections~\ref{sec:agent_roles}. Third, to automate fail2pass validation, we present our exit-code-based fail2pass validation in Section~\ref{sec:grading}. 


\subsection{Augmenting Raw Issue Data Collection to Include Binary Test Files}
\label{sec:binary_files_missing}

\begin{wrapfigure}{r}{0.5\textwidth}
    \vspace{-15pt} 
    \centering
    \includegraphics[width=1\linewidth]{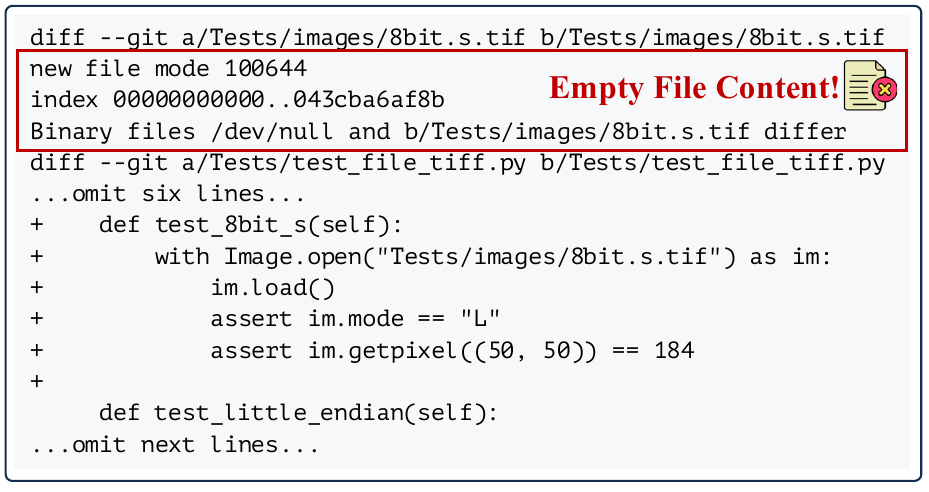} 
    \vspace{-20pt}
    \caption{Test patch of python-pillow\_\_Pillow-7111.}
    \label{fig:binary}
    \vspace{-10pt} 
\end{wrapfigure}

Following the methodology of SWE-bench, existing works typically rely on GitHub APIs and predefined patterns to collect issue–PR pairs for constructing issue resolution datasets. However, when a pull request involves binary files modification (e.g., .png, .tif, .zip), the patch directly downloaded via the GitHub API does not contain the actual binary content. As illustrated in Fig.~\ref{fig:binary}, the test patch only includes a placeholder message such as “Binary files ... differ”, leaving the file body empty. We term this phenomenon the Binary Test Files Missing Issue. To the best of our knowledge, we are the first to identify this problem in the existing pipeline.


This issue has two major consequences. We present an example\footnote{https://github.com/python-pillow/Pillow/pull/7111.diff} in Fig.~\ref{fig:binary}. First, it causes the target test cases to lack essential inputs.  We can find that the ``8bit.s.tif'' is directly used as the test input of the test function ``test\_8bit\_s''. When their contents are missing, the corresponding test code cannot obtain the required inputs and thus fails during execution, even if the test patch is applied successfully. Second, it leads to the failure of applying the test patch itself. Since the binary file is empty, the git apply operation cannot be completed, which also prevents subsequent modifications (e.g., changes in test\_file\_tiff.py) from being incorporated. As a result, the evaluation environment remains stuck at the base commit, without the intended test configuration. Consequently, these instances cannot pass the fail-to-pass validation step and are silently excluded from the final dataset.
We also find that this problem is particularly frequent in libraries for image processing and visualization, such as Pillow (30.43\% of instances affected), Manim (27.34\%), and matplotlib (22.10\%). This issue can limit the diversity of the collected data, especially for tasks that may require multi-modal understanding.

To address this problem, we adopt a simple yet effective strategy. We first use predefined patterns to detect such binary files and automatically generate direct download commands (e.g., via wget) to recover their contents. Then, we clean the test patch by removing incomplete binary-related hunks, ensuring that the remaining patch can be applied successfully. Although lightweight, this approach proves highly effective in Section~\ref{sec:ablation}, enabling us to collect task instances that would otherwise be discarded by existing methods.

\begin{figure}[t]
    \centering
    \includegraphics[width=0.95\linewidth]{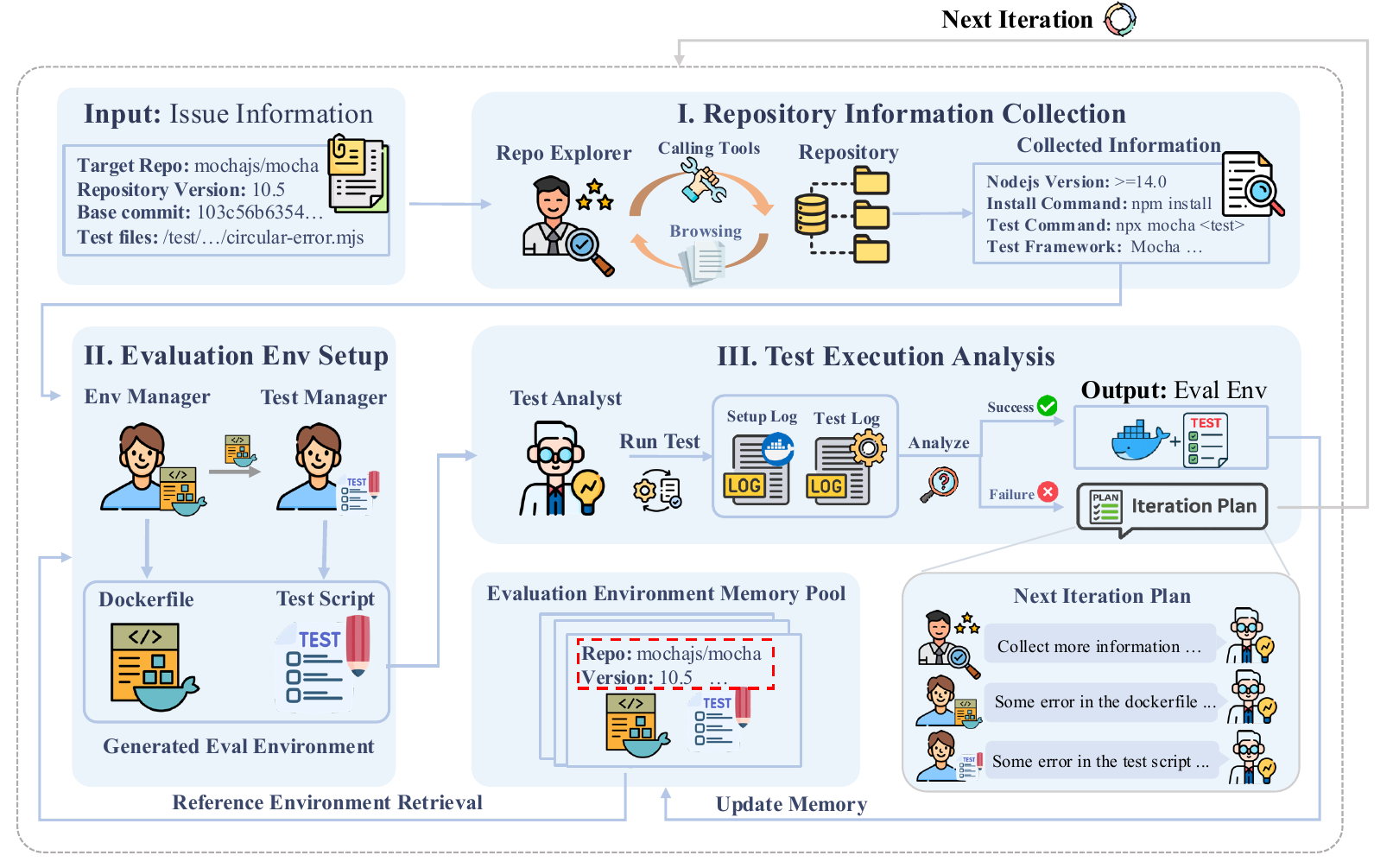}
    \caption{Framework overview of \setupagent.}
    \label{fig:overview}
\end{figure}

\begin{table}[h]
\centering
\caption{List of tools used in \setupagent.}
\label{tab:tools}
\small
\renewcommand{\arraystretch}{1.0}
\begin{tabular}{p{0.21\linewidth} p{0.48\linewidth} p{0.24\linewidth}}
\hline
\textbf{Tool Signature} & \textbf{Description} & \textbf{Output} \\
\hline
\texttt{browse\_file(fp, q)} & Extract content from file path \texttt{fp} relevant to query \texttt{q}. & Relevant snippets. \\
\hline
\texttt{browse\_dir(fp, d)} & Return directory tree of \texttt{fp} up to depth \texttt{d}. & Directory structure. \\
\hline
\texttt{search\_keyword(kw)} & Find file paths containing keyword \texttt{kw}. & Matched file paths. \\
\hline
\texttt{build\_image(df)} & Build Docker image from \texttt{df} (Dockerfile). & Image ID or build logs. \\
\hline
\texttt{start\_container(img)} & Launch container from image \texttt{img}. & Container ID or error logs. \\
\hline
\texttt{run\_eval(ct, es)} & Run evaluation script \texttt{es} in container \texttt{ct}. & Evaluation logs. \\
\hline
\end{tabular}
\end{table}

\subsection{Automatic Environemnt Setup Using \setupagent}
\label{sec:agent_roles}Inspired by success of previous works~\cite{liu2024marscode,gao2025trae,xia2024agentless,bouzenia2024repairagent,zhang2024autocoderover,ruan2024specrover,yang2024swe,yang2025enhancing,chen2025prometheus,wang2024aegis,ma2025thinking,jiang2025cosil,lin2024llms} in using LLM-based agent to resolve software engineering tasks automatically,  we propose \textbf{SWE-Builder}, a LLM-based multi-agent system to automate the evaluation environment construction for issues. We first introduce the roles of the four collaborative LLM-based agents designed to simulate the data construction process: (1) the Repository Explorer in Section~\ref{sec:role_repo}, (2) the Environment Manager in Section~\ref{sec:role_env}, (3) the Test Manager in Section~\ref{sec:role_eval_script}, and (4) the Test Analyst in Section~\ref{sec:role_test}. Following that, we describe SWE-Builder's memory mechanism to reuse previous experience, the Evaluation Environment Memory Pool, in Section~\ref{sec:memory}. Finally, we detail the complete workflow of SWE-Builder in Section~\ref{sec:agent_flow}. The overview of SWE-Builder overview is presented in Figure~\ref{fig:overview}. We provide the prompts for all proposed agents in our online Appendix\footnote{https://github.com/DeepSoftwareAnalytics/swe-factory/Appendix.pdf}.

\noindent
\begin{minipage}[t]{0.48\textwidth}
\subsubsection{Repository Explorer}
\label{sec:role_repo}
The repository explorer is responsible for automatically collecting the information essential for constructing an evaluation environment for each issue. For each target repository, it autonomously extracts (1) environment dependencies from configuration files (e.g., \texttt{requirements.txt} for Python, \texttt{pom.xml} for Java); (2) the relevant test commands (such as \texttt{pytest} for Python, \texttt{mvn test} for Java); and (3) additional setup details from documentation files (e.g., \texttt{CONTRIBUTING.md}) or installation scripts. By automating this process, the Repository Explorer efficiently handles the environment diversity across different projects without manual intervention.

\vspace{0.5em}
Algorithm~\ref{alg:context_retrieval} demonstrates the workflow of repository explorer. This agent begins by initializing its context with a task-specific prompt.
\end{minipage}
\hfill
\begin{minipage}[t]{0.49\textwidth}
\begin{algorithm}[H]
\caption{The workflow of Repository Explorer}
\label{alg:context_retrieval}
\KwIn{Repository $R$, LLM $\mathcal{L}$, Max Rounds $N$, Initial Context $\mathcal{C}$ ,}
\KwOut{Summary of setup information $S$ }
Initialize: $S \gets$ None\;
\For{$i = 1$ \KwTo $N$}{
    $actions, status, summary \gets \mathcal{L}(\mathcal{C})$\;
    \If{$status$ is true}{
        $S \gets summary$\;
        \Return{$S$}
    }
    \For{each API call $a_j$ in $actions$}{
        $o_j \gets$ Execute $a_j$ on $R$\;
        Update $\mathcal{C}$ with $o_j$\;
    }
    $S \gets summary$\;
}
\Return{$S$}
\end{algorithm}

\end{minipage}
\vspace{0.1em}
In each iteration, the agent queries the LLM with the  current context to decide which tools to use and whether sufficient setup information has been collected. Three tools from Table~\ref{tab:tools} are available: \texttt{browse\_file(fp, q)}, which calls the LLM to extract information relevant to the custom query from a given file; \texttt{browse\_directory(fp, d)}, which returns the structure of the specified directory up to a given depth; and \texttt{search\_keyword(kw)}, which retrieves file paths containing the target keyword. After executing the selected API calls and updating the context with the new observations, the agent summarizes the collected information. This process repeats until either the agent determines all necessary setup information has been obtained or the maximum number of rounds is reached, at which point the most recent summary is returned.

\subsubsection{Environment Manager}
\label{sec:role_env}
The environment manager is responsible for constructing a reliable runtime environment that ensures all tests related to the target issue can be executed correctly. To achieve this, the agent outputs a \texttt{Dockerfile}, which scripts the complex environment configuration and installation commands into a reproducible and portable format. The Environment Manager operates in a stepwise manner: after receiving comprehensive environment information from the Repo Explorer, it automatically generates or updates the \texttt{Dockerfile} to match the requirements of the repository and the specific issue. Importantly, the agent preserves its generation history across iterations, ensuring that all previous modifications are retained for robust and reproducible environment setup. In cases where generation fails, the previous \texttt{Dockerfile} is used as a fallback to guarantee continuity.

\subsubsection{Test Manager}
\label{sec:agent_test_generation}
\label{sec:role_eval_script}
The primary responsibility of the test manager is to automatically generate a shell script designed to execute tests relevant to a given issue. Its workflow commences only after its prerequisite agents—the Repository Explorer and Environment Manager—have completed their tasks. Leveraging the repository setup information from the Repository Explorer and the finalized Dockerfile from the Environment Manager, the test manager crafts the script to run within the specified containerized environment. Like the environment manager, it maintains a persistent context across iterations, ensuring that all script generation history is preserved.

Furthermore, to support other components in Section~\ref{sec:binary_files_missing} and Section~\ref{sec:grading}, the Test Manager's prompt includes specific instructions. First, to ensure binary file changes from test patches are correctly applied, the agent is prompted to insert the necessary commands into the test script, such as downloading new or modified files and deleting removed ones. Second, to ensure each test script provides a standardized interface for its final status, the agent is instructed to append commands that explicitly report the test's exit code. Following the convention where zero indicates success, the script captures the exit code of the main test command using \texttt{rc=\$?} and then prints a unique marker: \texttt{echo "OMNIGRIL\_EXIT\_CODE=\$rc"}. This allows other components to reliably and automatically parse the test status.

\subsubsection{Test Analyst}
\label{sec:role_test}
The test analyst evaluates the quality of the generated evaluation environment (including the Dockerfile and evaluation script), and based on the results, orchestrates multi-agent iterations to refine it. Its workflow begins after it receives these outputs from the Environment Manager and Test Manager. The agent operates on the fundamental assumption that a well-constructed environment must allow the issue's ground-truth patch to pass all relevant tests. To verify this, the agent follows a standardized procedure using a sequence of tools in Table \ref{tab:tools}. It first builds the Docker image with \texttt{build\_image(df)}, launches a container with \texttt{start\_container(img)}, and finally executes the evaluation script with \texttt{run\_eval(ct, es)} to apply the ground-truth patch and validate the tests. If the environment is successfully built, the agent analyzes the test logs to determine the test status. If the tests pass, the task is complete. If the tests fail, it analyzes the error information to generate an optimization plan. Similarly, if the environment construction fails, the agent analyzes the corresponding image build logs. In case of any failure, it generates a targeted plan and dispatches the relevant agents to initiate the next optimization iteration.

\subsubsection{ Evaluation Environment Memory Pool}
\label{sec:memory}
 We propose the evaluation environment memory pool, a component designed for \setupagent to reuse previously successful environment setups. This approach is motivated by a key observation: for issues within the same code repository, their required dependency environments and testing frameworks are often highly similar, especially for those from nearby versions. Building an evaluation environment from scratch for each issue is, therefore, both inefficient and prone to inconsistency. The memory pool addresses this challenge by archiving every successfully validated evaluation configuration, including both the dockerfile and the test script. Subsequently, when the Environment Manager and Test Manager handle a new issue, they first query the pool to find setups from the same repository and, from these, retrieve a reference environment setup from a nearby software version to use as a baseline. This strategy of reusing a pre-existing setup aims to accelerate the generation process and improve consistency across evaluation environments.

\subsubsection{Orchestration of Different Agents }
\label{sec:agent_flow}

The workflow of \setupagent is an iterative process that orchestrates the four agents to construct and refine a valid evaluation environment, as detailed in Algorithm~\ref{alg:swe_collector}. The process begins with a comprehensive initial iteration. In this first step, the Repository Explorer collects setup information while the system retrieves a relevant reference from the Evaluation Environment Memory Pool. This combined information serves as the initial context for the Environment Manager and Test Manager to generate the first versions of the Dockerfile and evaluation script, which are then passed to the Test Analyst for validation.

In contrast, subsequent iterations are not exhaustive but are instead targeted based on feedback from the Test Analyst. If a validation attempt fails, the Test Analyst identifies the cause of the error and provides specific guidance to the agent responsible for the flawed component. For example, if only the test script contains an error, the workflow invokes only the Test Manager to generate a revised script, bypassing the other agents. The corrected script is then passed back for another round of validation. This refinement cycle continues until the environment is successfully validated or the maximum number of iterations is reached.

\begin{algorithm}[ht]
\small
\caption{Workflow of \setupagent}
\label{alg:swe_collector}
\KwIn{
    Issue $\mathcal{I}$; evaluation environment memory pool $\mathcal{M}$; Max iteration number $\mathcal{N}$;\\
    Repo Explorer $\mathrm{Agent}_{\text{repo}}$; Environment Manager $\mathrm{Agent}_{\text{docker}}$; \\
    Test Manager $\mathrm{Agent}_{\text{script}}$;Test Analyst $\mathrm{Agent}_{\text{analyst}}$;
}
\KwOut{Dockerfile $\mathcal{D}$, evaluation script $\mathcal{S}$}

\textbf{Initialization:} $S_{\text{repo}}, S_{\text{dockerfile}}, S_{\text{script}} \gets \textbf{False}$;\\
\hspace*{1em} // $S_{\text{repo}}$: collected information ready; $S_{\text{dockerfile}}$: Dockerfile generated; $S_{\text{script}}$: test script generated\\ 

\For{$i = 1$ \KwTo $\mathcal{N}$}{
    \If{!$S_{\text{repo}}$}{
        collected\_information = $\mathrm{Agent}_{\text{repo}}$.run()\;
        \If{collected\_information}{
            $S_{\text{repo}} \gets \textbf{True}$;
            $\mathrm{Agent}_{\text{docker}}$.update\_context(collected\_information)\;
            $\mathrm{Agent}_{\text{script}}$.update\_context(collected\_information)\;
        }
    }
    reference\_val\_env = $\mathcal{M}$.retrieve\_closet\_version($\mathcal{I}$)\;
    $\mathrm{Agent}_{\text{docker}}$.update\_context(reference\_val\_env.dockerfile)\;
    $\mathrm{Agent}_{\text{script}}$.update\_context(reference\_val\_env.eval\_script)\;

    \If{$S_{\text{repo}}$ \textbf{and} !$S_{\text{dockerfile}}$}{
        $\mathcal{D}$, ok = $\mathrm{Agent}_{\text{docker}}$.run()\;
        \If{ok}{$S_{\text{dockerfile}} \gets \textbf{True}$;}
    }
    \If{$S_{\text{repo}}$ \textbf{and} $S_{\text{dockerfile}}$ \textbf{and} !$S_{\text{script}}$}{
        $\mathcal{S}$, ok = $\mathrm{Agent}_{\text{script}}$.run($\mathcal{D}$)\;
        \If{ok}{$S_{\text{script}} \gets \textbf{True}$;}
    }
    \If{$S_{\text{repo}}$ \textbf{and} $S_{\text{dockerfile}}$ \textbf{and} $S_{\text{script}}$}{
        analysis = $\mathrm{Agent}_{\text{analyst}}$.run($\mathcal{D}$, $\mathcal{S}$)\;
        \If{analysis.is\_finish}{
            $\mathcal{M}$.update($\mathcal{I}$, $\mathcal{D}$, $\mathcal{S}$)\;
            \Return $\mathcal{D}$, $\mathcal{S}$;
        }
        \If{analysis.guidance\_collected\_information}{
            $S_{\text{repo}} \gets \textbf{False}$; $\mathrm{Agent}_{\text{repo}}$.update\_context(analysis.guidance\_retrieval);}
        \If{analysis.guidance\_docker}{
            $S_{\text{dockerfile}} \gets \textbf{False}$; $\mathrm{Agent}_{\text{docker}}$.update\_context(analysis.guidance\_docker);}
        \If{analysis.guidance\_eval\_script}{
            $S_{\text{script}} \gets \textbf{False}$; $\mathrm{Agent}_{\text{script}}$.update\_context(analysis.guidance\_eval\_script);}
    }
}
\end{algorithm}


\subsection{Fail2Pass Validation Using Exit-Code-Based Test Log Parser}
\label{sec:grading}
Fail2pass validation is a critical step for ensuring the quality of benchmarks for the GitHub issue resolution task, following the methodology established by SWE-bench~\cite{jimenez2023swe}. After the successful construction of the evaluation environment for a given issue, the validation workflow for the instance involves three key stages: (1) applying test patche related to the given issue, and then executing the related tests on the codebase twice, once before and once after the gold patch is applied, to collect test logs from both runs; (2) manually inspecting these logs and developing custom parsers to extract the test status from them; and (3) retaining the instance for the final benchmark only if it exhibits a clear fail-to-pass transition.

However, the primary bottleneck in this phase is the manual process of inspecting logs and developing custom parsers to determine test status. This is a very challenging task because test log formats vary substantially across different programming languages, testing frameworks (e.g., pytest, JUnit, Jest), and repository configurations. Even within a single repository, testing framework versions and their output formats can change over time. Consequently, researchers often need to manually inspect a vast number of verbose test reports to understand their structure and then write repository-specific and version-specific parsing logic, typically relying on complex regular expressions. This process is very labor-intensive and severely limits the scalability and efficiency of constructing large-scale GitHub issue resolution benchmarks.

\begin{figure}[t]
    \centering
    \includegraphics[width=\linewidth]{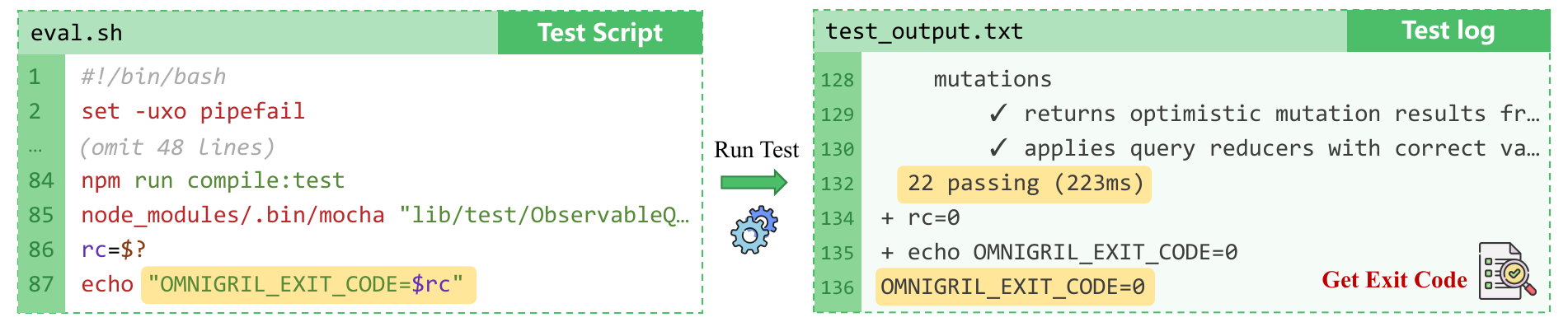}
    \caption{An example of capturing test command's exit code in script to reflect test status.}
    \label{fig:demo_exit_code}
\end{figure}

To automate the fail-to-pass validation pipeline, we replace complex log parsing with a standardized, exit-code-based approach. Our method leverages the common software engineering convention where a process exit code of zero signifies success and a non-zero value indicates failure. Instead of parsing diverse log formats, we modify the test execution script to report the outcome directly. This approach involves inserting two key commands into the test script immediately after the primary test command, as shown in the example in Figure~\ref{fig:demo_exit_code}. The first, \texttt{rc=\$?}, captures the exit code of the test process. The second, \texttt{echo "OMNIGRIL\_EXIT\_CODE=\$rc"}, prints this code prefixed with a unique, searchable identifier. This design establishes a simple and reliable interface for validation: the test status is determined by using a regular expression to match the OMNIGRIL\_EXIT\_CODE pattern, where an exit code of 0 indicates success and any non-zero value indicates failure. We implement this by providing a specific instruction in the prompt for our Test Manager agent (detailed in Section~\ref{sec:agent_test_generation}), which directs it to automatically include these commands to capture and report the exit code in the test scripts it generates.

The validation workflow using this exit-code approach is straightforward. After executing the test script for a given issue and obtaining the test log, our system parses the log to find the unique identifier, \texttt{OMNIGRIL\_EXIT\_CODE}. If the pattern is found, the captured exit code determines the outcome: a value of 0 signifies a successful test run (pass), while any non-zero value indicates a failure. If the identifier is not present in the log,  the status of test is marked with an error. For an instance to be considered valid, it must demonstrate a clear fail-to-pass transition, yielding a non-zero exit code before the gold patch is applied and a zero exit code after.

\section{Evaluation Setup}

In this paper, we evaluate \method from two aspects. First, we evaluate the effectiveness of \method in Section~\ref{sec:eval}, and the experimental setup is detailed in Section~\ref{sec:method_quality_setting}. Furthermore, we investigate whether \method can enhance a model's software engineering capabilities in Section~\ref{sec:exploratory_exp}, with the corresponding setup described in Section~\ref{sec:agent_setting}. 

\subsection{Experiment Details}
\label{sec:method_quality_setting}
\quad \textbf{Base Model Selection.} We select three models : \textit{GPT-4.1-mini-2025-04-14}~\cite{openai_gpt_4_1_mini}, \textit{\Kimi}~\cite{kimi_k2_2025}, and \textit{DeepSeek-V3-0324}~\cite{deepseek_v3_2024} as base models of \setupagent. Considering the high cost of running LLM-based agents, we do not use some of the most advanced models such as GPT-4.1 and Claude-4. The statistics of the selected base models are shown in Table~\ref{tab:model_stats}.

\begin{table}[t]
\footnotesize
\centering
\caption{Statistics of selected base models.  }
\label{tab:model_stats}
\begin{tabular}{lrrr}
\toprule
\textbf{Model} & \textbf{Input Cost} & \textbf{Output Cost} & \textbf{Release Date} \\
\midrule
\textit{GPT-4.1-mini-2025-04-14} & \$0.40 / 1M tokens & \$1.60 / 1M tokens & April 14, 2025 \\
\textit{\Kimi} & \$0.60 / 1M tokens & \$2.50 / 1M tokens & July 11, 2025 \\
\textit{DeepSeek-V3-0324} & \$0.27 / 1M tokens & \$1.10 / 1M tokens & March 24, 2025 \\
\bottomrule
\end{tabular}
\end{table}

\textbf{Evaluation Dataset Construction.} To evaluate the effectiveness of \method, we construct \textbf{SweSetupBench}, a dataset containing 671 raw issues data from 12 well-recognized open-source repositories spanning four programming languages. First,  we build a dataset of 2,441 issues from 12 open-source repositories. All selected repositories are well-recognized open-source projects, each with over 2.5k GitHub stars. These repositories span four programming languages—Python, Java, TypeScript, and JavaScript—which are among the most popular languages according to GitHub statistics. All issues included in the dataset are created before March 1st, 2025. Given the substantial computational cost of running LLM-based agent tools, we construct a smaller dataset for evaluation, refered to as SweSetupBench, by performing stratified sampling based on the repository version. Specifically, we randomly sample 20\% of the issues from each version. For versions with very few issues, we ensure that at least one issue is included to maintain comprehensive coverage. As a result, SweSetupBench contains 671 issues from 12 repositories across four languages. The detailed statistics of SweSetupBench are presented in Table~\ref{tab:swesetupbench_repo_stats}.

\begin{table}[t]
\footnotesize
\centering
\caption{Task instances statistics of SweSetupBench.}
\label{tab:swesetupbench_repo_stats}
\begin{tabular}{l l r r c r}
\toprule
\textbf{Repository Name} & \textbf{Language} & \textbf{\# Instances} & \textbf{\# Versions} & \textbf{Time Span} & \textbf{\# Stars} \\
\midrule
pallets/click                   & Python     & 32  & 11 & 2014--2025 & 16.5k \\
python-attrs/attrs              & Python     & 34  & 24 & 2016--2025 & 5.5k  \\
python-pillow/Pillow            & Python     & 132 & 48 & 2013--2025 & 12.8k \\
assertj/assertj                 & Java       & 39  & 32 & 2013--2025 & 2.7k  \\
checkstyle/checkstyle           & Java       & 77  & 49 & 2015--2025 & 8.6k  \\
eclipse-vertx/vert.x            & Java       & 70  & 14 & 2016--2024 & 14.5k \\
mochajs/mocha                   & JavaScript & 60  & 45 & 2012--2025 & 22.8k \\
iamkun/dayjs                    & JavaScript & 28  & 7  & 2018--2023 & 47.9k \\
nodejs/undici                   & JavaScript & 23  & 17 & 2024--2025 & 6.9k  \\
apollographql/apollo-client     & TypeScript & 78  & 29 & 2016--2025 & 19.6k \\
tailwindlabs/tailwindcss        & TypeScript & 72  & 25 & 2017--2025 & 88.3 k\\
reduxjs/redux-toolkit           & TypeScript & 26  & 13 & 2021--2025 & 11.0k \\
\bottomrule
\end{tabular}
\end{table}


\textbf{Hyperparameter Settings.} In our experiments, we set the maximum number of iterations for \setupagent to 5. The temperature of the base models is set to 0.1. We also set the maximum number of retrieval rounds for the context retrieval agent to 10. In addition, we run the experiments with 20 parallel subprocesses to improve efficiency. We empirically set these hyperparameters based on our preliminary experiments.

\textbf{Evaluation Metrics.} In evaluation, we use evaluation metrics as follows:

\begin{itemize}[leftmargin=15pt]
  \item Output Rate (\%): The proportion of tasks where our method successfully outputs a result.
\item F2P Rate (\%): The proportion of instances for which manual inspection confirms that the generated instance passes fail-to-pass validation. This process involves three experienced researchers manually labeling the outcome of each test log as ``fail'' or ``pass'' and resolving any disagreements through discussion.
  \item Time (minute): The average time required to process each task.
  \item Cost (\$): The average LLM API cost per task.
\end{itemize}

\subsection{Exploratory Experiment Details}
\label{sec:agent_setting}
\quad\textbf{Base Models for Training.} For the choice of base models for training, we select three code-specific LLMs: Qwen-2.5-Coder-instruct-3B, 7B, and 14B~\cite{qwen_qwen2_5_coder_2024}, as well as two general LLMs: Llama-3.1-8B-instruct~\cite{meta_llama_3_1} and Qwen3-8B-instruct~\cite{qwen3_2025}. Considering that Qwen3 features a Hybrid Thinking mode and our collected agent trajectories do not contain thinking mode information, we switch Qwen3 to a non-thinking mode for the subsequent training and inference phases.

\textbf{Training Details.} We train our models on a cluster of 8 A800 GPUs. We utilize the MS-Swift framework\footnote{https://github.com/modelscope/ms-swift} to perform full-parameter supervised fine-tuning (SFT) on the base models for 3 iterations. The training is conducted with a maximum sequence length of 65,536, bf16 precision, a learning rate of 1e-5, and a warmup ratio of 0.05. To enhance training efficiency, we enable FlashAttention~\cite{dao2022flashattention}. Furthermore, to optimize for long sequence training, we activate packing with the YaRN~\cite{peng2023yarn} method and employ sequence parallelism, setting sequence parallel size to 2 for the Qwen2.5-Coder-3B-instruct, 8 for the Qwen2.5-Coder-14B-instruct, and 4 for the other three LLMs.

\textbf{Evaluation Settings.} We evaluate performances of our trained models on a single A100 GPU. We use vLLM as the inference engine, with a maximum sequence length set to 32,768, bf16 precision, and prefix caching enabled. We use the DeepSWE agent~\cite{luo2025deepswe} as the agent framework. This agent framework equips the model with four custom-define tools to interact with the code environment: (1) \textit{Execute Bash} for running shell commands and capturing their output; (2) \textit{Search} for locating specific code snippets within files or directories; (3) \textit{File Editor} for viewing, creating, and modifying file contents; and (4) \textit{Finish/Submit} to signal the successful completion of the task. When running this agent, we set the model's temperature to 0 and the maximum number of agent iterations to 40. 

\textbf{Benchmarks and Metrics.} To evaluate model performance, we use two popular GitHub issue resolution benchmarks: SWE-bench-verified~\cite{openai2024swe} and SWE-bench-lite~\cite{jimenez2023swe}, containing 500 and 300 task instances, respectively. We use the following four metrics:

\begin{itemize}[leftmargin=15pt]
\item \textbf{Resolve Rate (\%):} the proportion of issues successfully resolved by the model.
\item \textbf{Empty Patch Rate (\%):}  the proportion of tasks where the model submits an empty patch.
\item \textbf{Tool Call Failure Rate (\%):} the rate of failed tool calls made by the model.
\item \textbf{Turns:} the average number of interaction rounds with the environment per task.
\end{itemize}

\section{Evaluation}
\label{sec:eval}
We summarize the following research questions (RQs) to evaluate \method:
\begin{itemize}
    \item \textbf{RQ1: What is  the Effectiveness of \setupagent?}
    \item \textbf{RQ2: How Much Do Different Components of \setupagent Contribute?}
    \item \textbf{RQ3: What is the Correctness of Exit-Code-Based Fail2pass Validation?}
\end{itemize}

\subsection{RQ1: Effectiveness of \setupagent}
\label{sec:effectiveness}
In this section, we evaluate the effectiveness of \setupagent on the SweSetupBench dataset.  We focus on two primary metrics: \textbf{Output Rate} and \textbf{Fail-to-Pass (F2P) Rate}. The Output Rate measures the proportion of tasks for which \setupagent successfully generates a complete evaluation environment (i.e., a Dockerfile and a test script). The F2P Rate, following the strict criteria of SWE-bench, measures the percentage of tasks that are manually verified to correctly transition from ``fail'' to ``pass'' after applying the gold patch, representing the portion of high-quality and valid instances for benchmarking.


\begin{table}[t]
  \centering
  \small
  \setlength{\tabcolsep}{5pt}
  \renewcommand{\arraystretch}{1.18}
  \caption{\setupagent 's performance on the SweSetupBench, with three different models. In this table,  GPT-4.1-mini  is short for GPT-4.1-mini-2025-04-14,  DeepSeek-V3 is short for DeepSeek-V3-0324, and ``min'' is short for minute. $\uparrow$ means higher is better, $\downarrow$ means lower is better. }
  \label{tab:models_perf_compact}
  \resizebox{0.85\textwidth}{!}{ 
  \begin{tabular}{llccccc}
    \toprule
    \textbf{Model}
      & \textbf{Dataset}
      & \textbf{F2P Rate (\%)}$\uparrow$
      & \textbf{Output Rate (\%)}$\uparrow$
      & \textbf{Time (min)}$\downarrow$
      & \textbf{Cost (\$)}$\downarrow$
      \\
    \midrule
    \multirow{5}{*}{\textbf{GPT-4.1 mini}}
      & Full    & \textbf{50.2} \textcolor{gray}{(337/671)}    & \textbf{64.8} \textcolor{gray}{(435/671)}   & 26.3   & 0.047   \\
      & Python  & 54.0 \textcolor{gray}{(107/198)}             &  \textbf{73.7} \textcolor{gray}{(146/198)}   &  19.0   & 0.040  \\
      & Java    & \textbf{43.5} \textcolor{gray}{(\phantom{1}81/186)}     &  \textbf{50.5} \textcolor{gray}{(\phantom{1}94/186)}   &  31.8  & 0.061  \\
      & TS      & 55.1 \textcolor{gray}{(\phantom{1}97/176)}              &  \textbf{68.2} \textcolor{gray}{(120/176)}   &  29.6 & 0.042   \\
      & JS      & \textbf{46.8} \textcolor{gray}{(\phantom{1}52/111)}              & \textbf{67.6} \textcolor{gray}{(\phantom{1}75/111)}   & 25.2   & 0.042  \\
    \midrule
    \multirow{5}{*}{\textbf{\Kimi}}
      & Full    & 47.8 \textcolor{gray}{(321/671)}             & 63.2 \textcolor{gray}{(424/671)}   & 30.2   & 0.056  \\
      & Python  & 54.0 \textcolor{gray}{(107/198)}             & 70.7 \textcolor{gray}{(140/198)}   &  26.5   &  0.051  \\
      & Java    & 33.9 \textcolor{gray}{(\phantom{1}63/186)}              & 48.4 \textcolor{gray} {(\phantom{1}90/186)}   &  38.9   &  0.073  \\
      & TS      & \textbf{56.3} \textcolor{gray}{(\phantom{1}99/176)}    & 68.2 \textcolor{gray}{(120/176)}   &  27.0   &  0.047   \\
      & JS      & \textbf{46.8} \textcolor{gray}{(\phantom{1}52/111)}     & 66.7 \textcolor{gray}{(\phantom{1}74/111)}   &  27.7   &  0.050 \\
    \midrule
    \multirow{5}{*}{\textbf{DeepSeek-V3}}
      & Full    & 42.0 \textcolor{gray}{(282/671)}             & 53.4 \textcolor{gray}{(358/671)}   & \textbf{23.0}   & \textbf{0.037}  \\
      & Python  & \textbf{54.5} \textcolor{gray}{(108/198)}    &  70.2 \textcolor{gray}{(139/198)}   &  16.2   &  0.030  \\
      & Java    & 30.1 \textcolor{gray}{(\phantom{1}56/186)}              &  39.8 \textcolor{gray}{(\phantom{1}74/186)}   &  26.4   &  0.047  \\
      & TS      & 42.0 \textcolor{gray}{(\phantom{1}74/176)}              &  49.4 \textcolor{gray}{(\phantom{1}87/176)}   &  28.6   &  0.035  \\
      & JS      & 39.6 \textcolor{gray}{(\phantom{1}44/111)}              &  52.3 \textcolor{gray}{(\phantom{1}58/111)}   &  20.9  & 0.035     \\
    \bottomrule
  \end{tabular}
  }
\end{table}
The experiment results, presented in Table~\ref{tab:models_perf_compact}, demonstrate that \setupagent is effective at creating evaluation environments. Overall,  all three models consistently achieve F2P rates over 42\% and Output Rates exceeding 53\%, with tasks completed in under 30 minutes on average and for less than \$0.06.  For example, with GPT-4.1 mini, \setupagent  achieves the best F2P Rate at 50.2\%, successfully producing 337 valid instances out of 671 issues, and also attains the highest Output Rate of 64.8\%. The performance of \Kimi is comparable, with a strong F2P Rate of 47.8\%, though at a slightly higher cost of \$0.056 per task. In contrast, DeepSeek-V3 stands out as the most cost-effective option, requiring the least amount of time (23.0 minutes) and the lowest cost (\$0.037) per task. In terms of efficiency, DeepSeek-V3 generates approximately 11.4 valid instances per dollar, significantly higher than GPT-4.1 mini (10.7) and \Kimi (8.5), making it a superior choice for large-scale processing under budget constraints.


We also observe that the performance of \setupagent varies across programming languages when paired with different models. For Python tasks, all three models achieve great and comparable F2P rates, with DeepSeek-V3 showing a slight advantage. On Java tasks, GPT-4.1 mini performs significantly better than the other two models. For TypeScript and JavaScript, both \Kimi and GPT-4.1 mini show strong and comparable results, outperforming DeepSeek-V3 by a notable margin.

\begin{center}
\begin{myboxc} \textbf{RQ1 Summary:} \setupagent can effectively construct valid evaluation environments at a reasonable cost for issues from repositories in different programming languages. \end{myboxc}
\end{center}

\subsection{RQ2: Ablation Studies of \setupagent}
\label{sec:ablation}
In this section,  we conduct a series of ablation studies to investigate the contribution of different components within \setupagent. These experiments target four key modules: (1) the Binary Test File Detecting method (Section~\ref{sec:binary_files_missing}), (2) the Repository Explorer (Section~\ref{sec:role_repo}), (3) the Evaluation Environment Memory Pool (Section~\ref{sec:memory}), and (4) Execution Feedback (Section~\ref{sec:role_test}). For the first three components, we directly remove these components to measure their impacts. For the fourth component, Execution Feedback, we alter the evaluation method. In our standard process, the Test Analyst provides guidance based on dynamic feedback from building the environment with the Dockerfile and executing the test script. For the ablation study, we replace this with a static analysis approach: the agent generates refinement suggestions by only inspecting the code of the Dockerfile and test script, without any actual execution.

First, We conduct an ablation study to evaluate the impact of our Binary Test File Detecting (BTFD) method. In SweSetupBench, there are 44 task instances where the test patch includes binary test files. To isolate the impact of our method, this study focuses exclusively on these 44 instances. As shown in Table \ref{tab:ablation_study_bftd}, removing the BTFD method (w/o BTFD) causes the F2P Rate of \setupagent to drop to 0\% across all models, with a general decline in the Output Rate as well. This is because the original test patch does not contain the actual content of the binary files, causing the test patch application to fail. Consequently, the codebase remains at the base commit, preventing the successful setup of a valid test environment. These results show that this component enables SWE-Builder to construct evaluation environments for issues that involve binary test files.

\begin{table}[t]
  \centering
  \footnotesize
  \setlength{\tabcolsep}{5pt}
  \renewcommand{\arraystretch}{1.18}
  \caption{Ablation study of Binary Test File Detecting method moudule in \setupagent. The BTFD is short for  Binary Test File Detecting method. $\uparrow$ means higher is better, $\downarrow$ means lower is better.}
  \label{tab:ablation_study_bftd}
  \resizebox{0.9\textwidth}{!}{ 
  \begin{tabular}{llcccc}
    \toprule
    \textbf{Model}
      & \textbf{Setting}
      & \textbf{F2P Rate (\%)}$\uparrow$
      
      & \textbf{Output Rate (\%)}$\uparrow$
      & \textbf{Time (min)}$\downarrow$
      & \textbf{Cost (\$)}$\downarrow$
      \\
    \midrule
    \multirow{2}{*}{\textbf{GPT-4.1 mini}}
      & Baseline    & 45.5 \textcolor{gray}{(\phantom{1}20/44)}       &  72.7 \textcolor{gray}{(32/44)}   &  22.0   & 0.042 \\
       & w/o BTFD  &  0 \textcolor{gray}{(\phantom{1}0/44)}       &  63.6 \textcolor{gray}{(28/44)}   & 23.6  &0.049  \\
    \midrule
    \multirow{2}{*}{\textbf{\Kimi}}
     & Baseline        &  50.0 \textcolor{gray}{(\phantom{1}22/44)}   &  63.6 \textcolor{gray}{(28/44)}   &  29.9   & 0.061  \\
       & w/o BTFD  &  0 \textcolor{gray}{(\phantom{1}0/44)}       &  59.1 \textcolor{gray}{(26/44)}   &  27.0   & 0.062  \\
    \midrule
    \multirow{2}{*}{\textbf{DeepSeek-V3}}
     & Baseline       &  50.0 \textcolor{gray}{(\phantom{1}22/44)}   &  50.0 \textcolor{gray}{(22/44)}   &   27.5  & 0.042  \\
       & w/o BTFD       &  0 \textcolor{gray}{(\phantom{1}0/44)}   &  50.0 \textcolor{gray}{(22/44)}   &  26.6 & 0.039     \\
    \bottomrule
  \end{tabular}
  }
\end{table}

We then evaluate the remaining three components on the full set of 671 issues in SweSetupBench, with results presented in Table \ref{tab:ablation_study}. First, removing the Repository Explorer leads to a decline in both F2P and Output rates for all models, demonstrating its positive contribution. The performance drop for \Kimi is less significant, and we believe this is because its training data may already contain information from these repositories. We also observe a substantial reduction in API costs with this removal, which we attribute to the elimination of frequent tool calls required for repository exploration. Second, removing the Memory Pool leads to an average decrease of 5.2\% in the F2P rate across all three models, demonstrating the value of reusing previous experiences. This removal also increases the average time, as more iterations are needed without prior experience to leverage. Third, removing the Execution Feedback component results in a significant performance collapse, with the F2P Rate decreasing to near zero. This highlights its critical role in guiding the construction of a valid environment. Although task time shortens by skipping the environment construction and test execution, costs increase because the models become uncertain and repeatedly attempt refinements, increasing the average number of iterations. Overall, these experiments confirm that all three components contribute effectively to the performance of \setupagent.

\begin{table}[t]
  \centering
  \footnotesize
  \setlength{\tabcolsep}{5pt}
  \renewcommand{\arraystretch}{1.18}
  \caption{Ablation studies of three components in \setupagent. Exec Feedback is short for Execution Feedback, Memory Pool is short for Evaluation Environment Memory pool. min is short for minute. DeepSeek-V3 is short for DeepSeek-V3-0324 $\uparrow$ means higher is better, $\downarrow$ means lower is better.}
  \label{tab:ablation_study}
  \resizebox{0.9\textwidth}{!}{ 
  \begin{tabular}{llcccc}
    \toprule
    \textbf{Model}
      & \textbf{Setting}
      & \textbf{F2P Rate (\%)}$\uparrow$
      & \textbf{Output Rate (\%)}$\uparrow$
      & \textbf{Time (min)}$\downarrow$
      & \textbf{Cost (\$)}$\downarrow$
      \\
    \midrule
    \multirow{4}{*}{\textbf{GPT-4.1 mini}}
      & Baseline    &  50.2 \textcolor{gray}{(337/671)}       & 64.8 \textcolor{gray}{(435/671)}   & 26.3   & 0.047 \\
        & w/o Repo Explorer  &  42.9 \textcolor{gray}{(288/671)}        &  55.3 \textcolor{gray}{(371/671)}   &  25.9   &  0.021 \\
      & w/o Memory Pool  &  45.0 \textcolor{gray}{(302/671)}     &   58.3 \textcolor{gray}{(391/671)}   &  30.0   & 0.050   \\
       & w/o Exec Feedback  &  0.3 \textcolor{gray}{(\phantom{1}2/671)}      &   3.8 \textcolor{gray}{(26/671)}   &  7.25   &  0.096  \\
    \midrule
    \multirow{4}{*}{\textbf{\Kimi}}
      & Baseline     & 47.8 \textcolor{gray}{(321/671)}      & 63.2 \textcolor{gray}{(424/671)}   & 30.2   & 0.056 \\
           & w/o Repo Explorer  &   47.1 \textcolor{gray}{(316/671)}       &  60.4 \textcolor{gray}{(405/671)}   &  32.6   &  0.032  \\
      & w/o Memory Pool  &  41.7 \textcolor{gray}{(280/671)}     &   54.7 \textcolor{gray}{(367/671)}   &  35.3   & 0.055  \\
       & w/o Exec Feedback  &  2.7 \textcolor{gray}{(\phantom{1}18/671)}      &   11.5 \textcolor{gray}{(77/671)} &6.1  &   0.077      \\
    \midrule
    \multirow{4}{*}{\textbf{DeepSeek-V3}}
      & Baseline       & 42.0 \textcolor{gray}{(282/671)}   & 53.4 \textcolor{gray}{(358/671)}   & 23.0   & 0.037  \\
           & w/o Repo Explorer  &   33.5 \textcolor{gray}{(225/671)}     &  44.3 \textcolor{gray}{(297/671)}   &  20.0   &  0.016   \\
      & w/o Memory Pool  &  37.7 \textcolor{gray}{(253/671)}      &   45.8 \textcolor{gray}{(307/671)}  & 27.5 &  0.038    \\
       & w/o Exec Feedback  & 0.7  \textcolor{gray}{(\phantom{1}5/671)}     & 16.4  \textcolor{gray}{(110/671)}   &   15.8  & 0.065   \\
    \bottomrule
  \end{tabular}
  }
\end{table}

\begin{center}
\begin{myboxc}
\textbf{RQ2 Summary:} All components contribute to the effectiveness of \setupagent. First, the Binary Test File Detecting method is essential for building test environments for issues that include binary files. Second, the Repository Explorer improves results by providing helpful context, but at the cost of more API calls. Third, the Evaluation Environment Memory Pool boosts performance and shortens run time by reusing past solutions. Finally, Execution Feedback is the most critical part. Without it, performance collapses and the F2P Rate drops to nearly zero, making it almost impossible to build the test environment, even though it runs faster.
\end{myboxc}
\end{center}




\subsection{RQ3: Correctness of Exit-Code-Based Fail2pass Validation}
\label{sec:rq3}
In this section, we conduct a human evaluation to investigate the correctness of our exit-code-based fail2pass validation method. First, we collect 1217 task instances generated from GPT-4.1 mini, \Kimi, and DeepSeek-V3-0324 as described in Section~\ref{sec:effectiveness}. For each instance, we execute the test scripts before and after applying the gold patch to obtain a pair of test logs. We successfully obtained test log pairs for 1201 instances, after excluding those that failed due to environment build errors or test timeouts. We then employ our exit-code-based log parser to automatically determine the test status. Following SWE-bench~\cite{jimenez2023swe}, if the status changes from ``fail'' to ``pass'', the instance is labeled as ``valid''; otherwise, it is labeled ``invalid''. Finally, to validate our method, we establish the ground truth through manual evaluation. Three experienced researchers manually inspect each test log, labeling its outcome as ``fail'' or ``pass''. If there are any disagreement in their labels, the three researchers will discuss to resolve them. Based on these consensus labels, we then determine the validity of each instance. We then compare the labels generated by our method against this ground-truth. The results are presented in Table~\ref{tab:manual_inspect}.

\begin{figure}[t]
    \centering
    \includegraphics[width=\linewidth]{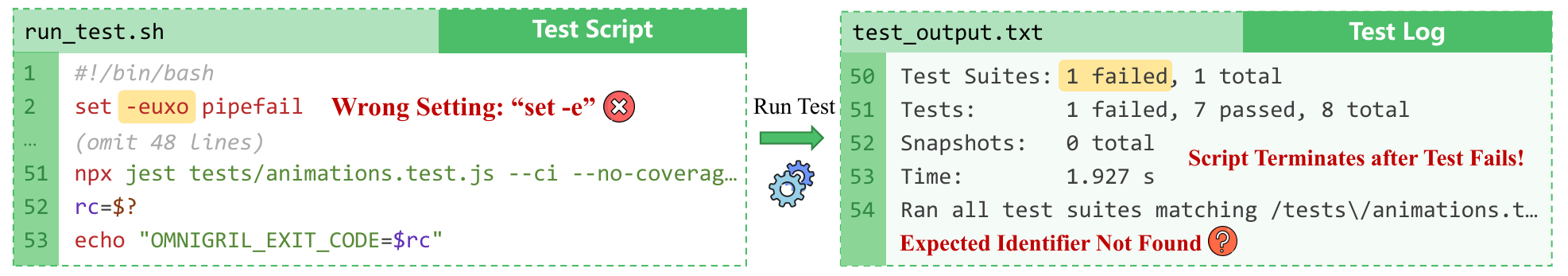}
    \caption{An example of early test script termination phenomenon.}
    \label{fig:wrong_category}
\end{figure}

From experiment results from Table~\ref{tab:manual_inspect}, we can find that our method shows a high accuracy in judging valid task instances, with a F1 score of 0.99. The small number of false negatives (FN) stem from 16 task instances that are classified as invalid due to the failure to parse a test status from their logs. Further analysis revealed that this is due to issues in the test scripts generated by LLM, which can be divided into two types: (1) Incorrect Identifier (4/16). This type means the test script fails to use our expected unique identifier (\texttt{OMNIGRIL\_EXIT\_CODE}), for example, using a typo like \texttt{OMNIGRIL\_CODE}, which prevents our parser from extracting the status. (2) Early Script Termination (12/16). As shown in Figure ~\ref{fig:wrong_category}, the test script incorrectly includes a ``set -e'' command (a shell option to exit immediately on error). This causes the test script to terminate upon test failure before it could execute the \texttt{echo ``OMNIGRIL\_EXIT\_CODE=\$rc''} line, thus preventing us from parsing the test status. We attribute these script generation errors to the stochastic nature of the LLM's inference, which sometimes does not guarantee strict adherence to instructions. Fortunately, the resulting issues can be reliably detected through simple pattern matching and corrected manually.

Furthermore, to investigate whether the exit code itself accurately reflects the test status, we examine the individual logs from the aforementioned fail-to-pass runs. From the instances without script errors, we collect all 2,380 test logs that correctly outputted an exit code, labeling logs with an exit code of 0 as ``pass'' and non-zero as ``fail''. Manual inspection confirm the accuracy of this method, which achieves a high F1 score of 1.0 and demonstrates its reliability to determine the status of individual tests.

\begin{table}[t]

  \centering
  \caption{Human evaluation results on the correctness of exit-code-based fail2pass validation method.}
  \label{tab:manual_inspect}
  \resizebox{0.9\textwidth}{!}{
  \begin{tabular}{lcccccccc}
    \toprule
    \textbf{Data Sources From} & \textbf{\# Task Instances} & \textbf{\# TP} & \textbf{\# FP} & \textbf{\# TN} & \textbf{\# FN} & \textbf{Precision (\%)} & \textbf{Recall (\%)} & \textbf{F1}\\
    \midrule
    GPT-4.1 mini & 434 & 337 & 0 & 97 & 0 & 100 & 100 & 1.000 \\
    DeepSeek-V3  & 352 & 277 & 0 & 71 & 4 & 100 & 98.6 & 0.993 \\
    \Kimi      & 415 & 309 & 0 & 94 & 12 & 100 & 96.3 & 0.981 \\
    \midrule
    \textbf{Total} & \textbf{1,201} & \textbf{923} & \textbf{0} & \textbf{262} & \textbf{16} & \textbf{100} & \textbf{98.3} & \textbf{0.991} \\
    \bottomrule
  \end{tabular}}
\end{table}

\begin{center}
\begin{myboxc} \textbf{RQ3 Summary:} Our human evaluation demonstrates the high accuracy of the exit-code-based fail2pass validation method. Besides, we also verify that the exit code is a reliable indicator for determining the test status. \end{myboxc}
\end{center}

\section{Exploratory Experiment: Enhance SE Ability of LLMs Using \method}
\label{sec:exploratory_exp}
 In this section, we explore whether \method can enhance the software engineering ability of models. Following previous studies~\cite{ma2025sorft,wei2025swe,zhang2025sealign,ma2024lingma}, we construct agent training data to improve models' performance. In Section~\ref{sec:agent_training}, we will describe details about how to construct agent training data using \method. And in Section~\ref{sec:agent_evaluation}, we will evaluate the performance of our trained model on two popular benchmarks: SWE-bench-verified~\cite{openai2024swe} and SWE-bench-lite~\cite{jimenez2023swe}. The detailed experiment settings are introduced in Section~\ref{sec:agent_setting}. 

\subsection{Agent Training Data Construction using \method} 
\label{sec:agent_training}
\quad \textbf{Training Task Instances Collection Using \method}. We select 10 popular Python repositories and leveraged \method to automatically construct 2877 task instances using GPT-4.1 mini. . To prevent data leakage during evaluation on SWE-bench, we ensure that the repositories selected for data construction are distinct from those in the SWE-bench benchmark. Each task instance includes basic issue information, the Docker-based evaluation environment, and associated test scripts. In our experiments, we employ the DeepSWE agent~\cite{luo2025deepswe} as our agent framework. To ensure the agent could interact correctly within the environment of each task instance, we used GPT-4.1 mini to install the four custom-defined tools the agent framework relies on. This process ultimately yielded 2,809 fully prepared agent training environments for training. The final dataset statistics is present in our online Appendix\footnote{https://github.com/DeepSoftwareAnalytics/swe-factory/Appendix.pdf}

\textbf{Agent Trajectory Collection for LLM Training.} Following previous studies~\cite{yang2025swe,pan2024training,jain2025r2e}, we collect agent trajectories—defined as the multi-turn interaction logs between a capable agent model and the environment—to train the model's agent capabilities. For this step, we select \Kimi as the agent model because of its strong performance on SWE-bench and its reasonable cost. Subsequently, we use the DeepSWE agent~\cite{luo2025deepswe} as the agent framework, setting the model's temperature to 0.2 and the maximum number of agent iterations to 40. For each of the 2,809 agent training environments we construct, we sample a single trajectory, resulting in a total of 2,809 trajectories for our agent training dataset for models.


\begin{table}[t]
\centering
\caption{Performances of base models and fintuned models on \textbf{SWE-Bench Verified} (top) and \textbf{SWE-Bench Lite} (bottom). 
Arrows indicate the preferred direction: $\uparrow$ means higher is better, $\downarrow$ means lower is better. 
Qwen2.5-Coder denotes \emph{Qwen2.5-Coder-Instruct}. 
EP is the Empty Patch Rate (\%), TF is the Tool Call Failure Rate (\%), RR is the Resolve Rate (\%), 
and ``Turns'' is the average number of interaction rounds.}
\setlength{\tabcolsep}{4pt}
\renewcommand{\arraystretch}{1.12}
\small
\resizebox{\linewidth}{!}{%
\begin{tabular}{l c
                 ccc
                 ccc
                 ccc
                 YYY}
\toprule
\multirow{2}{*}{Model} & \multirow{2}{*}{Size} &
\multicolumn{3}{c}{EP (\%,\,\(\downarrow\))} &
\multicolumn{3}{c}{TF (\%,\,\(\downarrow\))} &
\multicolumn{3}{c}{Turns} &
\multicolumn{3}{c}{\cellcolor{ResolveBG}RR (\%,\,\(\uparrow\))} \\
\cmidrule(lr){3-5}\cmidrule(lr){6-8}\cmidrule(lr){9-11}\cmidrule(lr){12-14}
& & Base & SFT & $\Delta$ & Base & SFT & $\Delta$ & Base & SFT & $\Delta$ 
& \cellcolor{ResolveBG}base & \cellcolor{ResolveBG}SFT & \cellcolor{ResolveBG}$\Delta$ \\

\midrule
\rowcolor[HTML]{F2F2F2} \multicolumn{14}{c}{\textbf{SWE-Bench Verified (500 instances)}} \\
Qwen2.5-Coder & 3B  & 90.8  & 16.4  & \good{$-74.4$} & 92.8  & 8.9  & \good{$-83.9$} & 28.10 & 39.26 & $+11.2$ & 0.0 & 3.4  & \good{$+3.4$} \\
Qwen2.5-Coder & 7B  & 91.0  & 4.4   & \good{$-86.6$} & 82.4  & 1.01 & \good{$-81.4$} & 15.02 & 38.91 & $+23.9$ & 0.0 & 13.6 & \good{$+13.6$} \\
Qwen2.5-Coder & 14B & 4.6   & 6.6   & \bad{$+2.0$}   & 5.1   & 0.5  & \good{$-4.6$}  & 37.22 & 37.50 & $+0.3$  & 5.8 & 21.0 & \good{$+15.2$} \\
Qwen-3-instruct        & 8B  & 78.0  & 17.4  & \good{$-60.6$} & 0.2   & 4.1  & \bad{$+3.9$}   & 38.64 & 39.33 & $+0.7$  & 3.4 & 16.2 & \good{$+12.8$} \\
Llama-3.1-instruct     & 8B  & 41.0  & 19.60 & \good{$-21.4$} & 43.2  & 5.6  & \good{$-37.6$} & 38.76 & 39.34 & $+0.6$  & 1.2 & 5.2  & \good{$+4.0$} \\
\midrule
\rowcolor[HTML]{F2F2F2} \multicolumn{14}{c}{\textbf{SWE-Bench Lite (300 instances)}} \\
Qwen2.5-Coder & 3B  & 94.0  & 14.0  & \good{$-80.0$} & 91.07 & 9.13 & \good{$-81.9$} & 27.74 & 39.34 & $+11.6$ & 0.0 & 3.3 & \good{$+3.3$} \\
Qwen2.5-Coder & 7B  & 94.33 & 3.3   & \good{$-91.0$} & 82.62 & 0.83 & \good{$-81.8$} & 13.72 & 38.70 & $+25.0$ & 0.0 & 8.0 & \good{$+8.0$} \\
Qwen2.5-Coder & 14B & 6.33  & 6.67  & \bad{$+0.34$}  & 5.80  & 0.47 & \good{$-5.3$}  & 36.59 & 37.62 & $+1.0$  & 4.3 & 14.7 & \good{$+10.4$} \\
Qwen-3-instruct        & 8B  & 78.67 & 19.33 & \good{$-59.3$} & 0.25  & 4.14 & \bad{$+3.9$}   & 38.68 & 39.40 & $+0.7$  & 3.0 & 11.3 & \good{$+8.3$} \\
Llama-3.1-instruct     & 8B  & 45.0  & 20.0  & \good{$-25.0$} & 43.32 & 5.46 & \good{$-37.9$} & 39.06 & 39.31 & $+0.3$  & 1.0 & 4.3  & \good{$+3.3$} \\
\bottomrule
\end{tabular}
}
\label{tab:swebench-combined}
\end{table}

\subsection{Performances of Trained Models}
\label{sec:agent_evaluation}
We evaluate the software engineering ability of the base models and the finetuned models on both SWE-bench-verified and SWE-bench-lite. The evaluation settings are detailed in Section~\ref{sec:agent_setting}.

From the results in Table~\ref{tab:swebench-combined}, we can find some key observations. First, all LLMs show an improved resolve rate on both datasets after fine-tuning, which demonstrates the effectiveness of the training data collected by our method. Second, we find that the performance improvement after fine-tuning is more significant for models with larger parameter sizes. We attribute this to the stronger foundational coding abilities of larger models, allowing finetuning to unlock greater potential. Third, we notice that finetuned models tend to have more interaction turns with the environment, show a low empty patch rate, and show a lower tool call failure rate. We believe that SFT improves model performance by enhancing its capabilities in multi-turn interaction, code editing, and tool calling. Overall, our experiments demonstrate the potential of \method for enhancing the software engineering capabilities of models.

\begin{center}
\begin{myboxc} \textbf{Exploratory Experiment Summary:} \method can effectively enhance the software engineering ability of models by constructing training datasets automatically. \end{myboxc}
\end{center}

\section{Related Work}
\subsection{Datasets for GitHub Issue Resolution}

Recently, many benchmarks have been proposed to evaluate the ability of large language models (LLMs) to resolve real-world GitHub issues. Among them, SWE-bench~\cite{jimenez2023swe} is the most widely used, providing 2,294 Python issues with paired post-PR test suites for automated evaluation. Subsequently, some works such as SWE-bench-Verified~\cite{openai2024swe} improved SWE-bench to enhance its reliability~\cite{openai2024swe,wang2025solved,huang2024agents,aleithan2024swe,yu2025utboost}. Following that, other works such as Multi-SWE-bench~\cite{zan2025multi} propose new issue resolution datasets covering more languages and modalities~\cite{guo2025omnigirl, yang2024swem, zan2025multi, zan2024swe, aleithan2024swe,deng2025nocode,zhang2025swef,du2025swe}. Besides, to improve the issue resolution ability of models by training, researchers propose a number of training datasets~\cite{pan2024training} that support large-scale training and automated evaluation. For example, SWE-Gym~\cite{pan2024training} provides 2,438 Python tasks with Docker-based environments and unit tests for agent training. Additionally, other researchers~\cite{jain2025r2e,yang2025swe,pham2025swe} use LLM to synthesize issue data to train the issue resolution ability of models. For example, R2E-Gym~\cite{jain2025r2e} uses LLM to generate 8,700 task instances using automated test synthesis and back-translation.


\subsection{Automatic Environment Setup}
Recently, many works~\cite{bouzenia2024you,eliseeva2025envbench,hu2025llm,yu2025cxxcrafter} try to use LLM to automate repository environment construction.  For example, ExecutionAgent~\cite{bouzenia2024you} is an LLM-based agent capable of constructing repository environments across many languages. However, while these works address general environment setup, their objectives and methods are not directly applicable to the construction of issue resolution datasets, which has more specialized requirements. Subsequently, some researchers~\cite{zhang2025swe,vergopoulos2025automated} have focused specifically on automated environment setup for the issue resolution task. For example, SetupAgent~\cite{vergopoulos2025automated} shows a strong ability to automate environment construction for issue resolution, but this work is not open-source. Compared to our method, these approaches still rely on manually written test log parsers, which require manual fail-to-pass (F2P) validation. Furthermore, these existing works are only applicable to Python repositories.

\section{Threats to Validity and Limitations}

A potential threat to validity is the scope of our evaluation dataset. To enhance the diversity of our data, we collect issues in four different programming languages from 12 repositories and ensure variety across the issue versions. In the future, we plan to conduct a more extensive evaluation on a more diverse dataset to enhance the generalizability of our conclusions. Another threat to validity is the selection of models for training in Section~\ref{sec:exploratory_exp}. To ensure model diversity, we select models with various parameter scales, from different model families and versions. However, due to computational resource constraints, our experiments are limited to models with up to 14B parameters. In the future, we plan to extend our experiments to a broader range of models.



\section{Conclusion}

In this paper, we propose \method, an automatic benchmark construction pipeline for the GitHub issue resolution task. First, we identify a binary test file missing issue in the current data construction pipeline and propose methods to resove it.  Second, we introduce a multi-agent system called \setupagent to automate evaluation environment construction with collaboration among LLM-based agents. To automate the fail2pass validation stage, we propose an exit-code-based log parsing method to robustly extract test status from test log without the need for manual parsing or inspection. Our experiments show that \setupagent can effectively construct evaluation environment for issues. For example, using GPT-4.1 mini, \method successfully constructs  evaluation environment for 50.2\% issues. And our ablation studies show that all components contribute to the effectiveness of \setupagent. Besides, our human evaluation show that our exit-code-based fail2pass method achieves a high accuracy. Finally, our exploratory demonstrates that \method can effectively improve the software engineering of models. We hope \method can accelerate the collection of large-scale, high-quality GitHub issue resolution datasets for training and evaluation.

\section{Data Availability}
\label{sec:open-source}
We have released our code, task instances, agent trajectories, and trained models at \url{https://github.com/DeepSoftwareAnalytics/swe-factory}.

\bibliographystyle{ACM-Reference-Format}
\bibliography{ref}
\end{document}

%% file: tool.tex
\usepackage{tcolorbox}
\newcommand{\method}{SWE-Factory\xspace}
\newcommand{\setupagent}{SWE-Builder\xspace}
\newcommand{\boxmargin}{1mm}
\newtcolorbox{myboxc}{
    colback=gray!15!white,
    arc = 0pt, outer arc = 0pt,
    boxsep=0pt, left = 3pt, right = 0pt, top = 0pt, bottom = 0pt, 
    leftrule=3pt, bottomrule=0pt,toprule=0pt, rightrule=0pt,
    left = \boxmargin, right = \boxmargin, top = \boxmargin, bottom = \boxmargin
}

\newcommand{\Kimi}{Kimi-K2\xspace}